\input harvmac.tex
\input epsf.tex

\def\figin{\epsfcheck\figin}\def\figins{\epsfcheck\figins}
\def\epsfcheck{\ifx\epsfbox\UnDeFiNeD
\message{(NO epsf.tex, FIGURES WILL BE IGNORED)}
\gdef\figin##1{\vskip2in}\gdef\figins##1{\hskip.5in}
\else\message{(FIGURES WILL BE INCLUDED)}%
\gdef\figin##1{##1}\gdef\figins##1{##1}\fi}
\def\DefWarn#1{}
\def\figinsert{\goodbreak\midinsert}
\def\ifig#1#2#3{\DefWarn#1\xdef#1{fig.~\the\figno}
\writedef{#1\leftbracket fig.\noexpand~\the\figno}%
\figinsert\figin{\centerline{#3}}\medskip\centerline{\vbox{\baselineskip12pt
\advance\hsize by -1truein\noindent\footnotefont{\bf
Fig.~\the\figno:} #2}}
\bigskip\endinsert\global\advance\figno by1}


\def\frac#1#2{{#1 \over #2}}
\def\text#1{#1}

\lref\kallosh{
  R.~Kallosh and A.~A.~Tseytlin,
  JHEP {\bf 9810}, 016 (1998)
  [arXiv:hep-th/9808088].
}

\lref\Siegel{
  W.~Siegel,
  Phys.\ Lett.\  B {\bf 84}, 193 (1979).
}
\lref\EllisQJ{
  R.~K.~Ellis, W.~J.~Stirling and B.~R.~Webber,
  ``QCD and collider physics,''
  Camb.\ Monogr.\ Part.\ Phys.\ Nucl.\ Phys.\ Cosmol.\  {\bf 8}, 1 (1996).
}

\lref\grossetal{
  D.~J.~Gross, A.~Hashimoto and N.~Itzhaki,
  Adv.\ Theor.\ Math.\ Phys.\  {\bf 4}, 893 (2000)
  [arXiv:hep-th/0008075].
}
\lref\ItzhakiDD{
  N.~Itzhaki, J.~M.~Maldacena, J.~Sonnenschein and S.~Yankielowicz,
  Phys.\ Rev.\  D {\bf 58}, 046004 (1998)
  [arXiv:hep-th/9802042].
}
 \lref\wilsonloop{   J.~M.~Maldacena,
  Phys.\ Rev.\ Lett.\  {\bf 80}, 4859 (1998)
  [arXiv:hep-th/9803002].
 S.~J.~Rey and J.~T.~Yee,
  Eur.\ Phys.\ J.\  C {\bf 22}, 379 (2001)
  [arXiv:hep-th/9803001].
  }

 \lref\adscft{J.~M.~Maldacena,
  Adv.\ Theor.\ Math.\ Phys.\  {\bf 2}, 231 (1998)
  [Int.\ J.\ Theor.\ Phys.\  {\bf 38}, 1113 (1999)]
  [arXiv:hep-th/9711200].
}

\lref\magoo{O.~Aharony, S.~S.~Gubser, J.~M.~Maldacena, H.~Ooguri
and Y.~Oz,
  Phys.\ Rept.\  {\bf 323}, 183 (2000)
  [arXiv:hep-th/9905111].
} \lref\zvi{ Bern et al,  0505205 }

\lref\mkru{ M.~Kruczenski, R.~C.~Myers and A.~W.~Peet,
  JHEP {\bf 0205}, 039 (2002)
  [arXiv:hep-th/0204144].
}

\lref\bernetal{  Z.~Bern, L.~J.~Dixon and V.~A.~Smirnov,
  Phys.\ Rev.\  D {\bf 72}, 085001 (2005)
  [arXiv:hep-th/0505205].
}

\lref\integrothers{ N.~Beisert, R.~Hernandez and E.~Lopez,
  JHEP {\bf 0611}, 070 (2006)
  [arXiv:hep-th/0609044].
 B.~Eden and M.~Staudacher,
  J.\ Stat.\ Mech.\  {\bf 0611}, P014 (2006)
  [arXiv:hep-th/0603157].
}

\lref\MagneaZB{
  L.~Magnea and G.~Sterman,
  Phys.\ Rev.\  D {\bf 42}, 4222 (1990).
}

\lref\BernKQ{
  Z.~Bern, L.~J.~Dixon and D.~A.~Kosower,
  Comptes Rendus Physique {\bf 5}, 955 (2004)
  [arXiv:hep-th/0410021].
}
\lref\StermanQN{
  G.~Sterman and M.~E.~Tejeda-Yeomans,
  Phys.\ Lett.\  B {\bf 552}, 48 (2003)
  [arXiv:hep-ph/0210130].
}

\lref\CataniBH{
  S.~Catani,
  Phys.\ Lett.\  B {\bf 427}, 161 (1998)
  [arXiv:hep-ph/9802439].
}

\lref\BrockSZ{
  R.~Brock {\it et al.}  [CTEQ Collaboration],
  ``Handbook of perturbative QCD: Version 1.0,''
  Rev.\ Mod.\ Phys.\  {\bf 67}, 157 (1995).
}

\lref\StermanWJ{
  G.~Sterman and S.~Weinberg,
  Phys.\ Rev.\ Lett.\  {\bf 39}, 1436 (1977).
}
\lref\BernEW{
  Z.~Bern, M.~Czakon, L.~J.~Dixon, D.~A.~Kosower and V.~A.~Smirnov,
  arXiv:hep-th/0610248.
}

\lref\RhoJM{
  M.~Rho, S.~J.~Sin and I.~Zahed,
  Phys.\ Lett.\  B {\bf 466}, 199 (1999)
  [arXiv:hep-th/9907126].
}

\lref\JanikZK{
  R.~A.~Janik and R.~Peschanski,
  Nucl.\ Phys.\  B {\bf 565}, 193 (2000)
  [arXiv:hep-th/9907177].
}

\lref\BrodskyPX{
  S.~J.~Brodsky and G.~F.~de Teramond,
  Phys.\ Lett.\  B {\bf 582}, 211 (2004)
  [arXiv:hep-th/0310227].
}

\lref\AndreevVU{
  O.~Andreev,
  Phys.\ Rev.\  D {\bf 70}, 027901 (2004)
  [arXiv:hep-th/0402017].
}

\lref\BrowerEA{
  R.~C.~Brower, J.~Polchinski, M.~J.~Strassler and C.~I.~Tan,
  arXiv:hep-th/0603115.
}

\lref\gkp{ S.~S.~Gubser, I.~R.~Klebanov and A.~M.~Polyakov,
  Nucl.\ Phys.\  B {\bf 636}, 99 (2002)
  [arXiv:hep-th/0204051].
}

\lref\bes{
  N.~Beisert, B.~Eden and M.~Staudacher,
  J.\ Stat.\ Mech.\  {\bf 0701}, P021 (2007)
  [arXiv:hep-th/0610251].
}

\lref\FrolovAV{
  S.~Frolov and A.~A.~Tseytlin,
  JHEP {\bf 0206}, 007 (2002)
  [arXiv:hep-th/0204226].
}

\lref\grossmende{  D.~J.~Gross and P.~F.~Mende,
  Phys.\ Lett.\  B {\bf 197}, 129 (1987).
D.~J.~Gross and P.~F.~Mende,
  Nucl.\ Phys.\  B {\bf 303}, 407 (1988).
}
\lref\spiky{
 M.~Kruczenski,
  JHEP {\bf 0508}, 014 (2005)
  [arXiv:hep-th/0410226].
}

\lref\polchstrass{
  J.~Polchinski and M.~J.~Strassler,
  Phys.\ Rev.\ Lett.\  {\bf 88}, 031601 (2002)
  [arXiv:hep-th/0109174].
 J.~Polchinski and M.~J.~Strassler,
  JHEP {\bf 0305}, 012 (2003)
  [arXiv:hep-th/0209211].
}

\lref\beisertsnlin{ N.~Beisert,
  J.\ Stat.\ Mech.\  {\bf 0701}, P017 (2007)
  [arXiv:nlin.si/0610017].
}

\lref\CachazoAZ{
  F.~Cachazo, M.~Spradlin and A.~Volovich,
  arXiv:hep-th/0612309.
}

\lref\kruczenski{
 M.~Kruczenski,
  JHEP {\bf 0212}, 024 (2002)
  [arXiv:hep-th/0210115].
  }

\lref\CollinsBT{
  J.~C.~Collins,
  Adv.\ Ser.\ Direct.\ High Energy Phys.\  {\bf 5}, 573 (1989)
  [arXiv:hep-ph/0312336].
}

\lref\KorchemskySI{
  G.~P.~Korchemsky,
  Mod.\ Phys.\ Lett.\  A {\bf 4}, 1257 (1989).
}

\lref\GrisaruVM{
  M.~T.~Grisaru, H.~N.~Pendleton and P.~van Nieuwenhuizen,
  Phys.\ Rev.\  D {\bf 15}, 996 (1977).
}

\lref\GrisaruPX{
  M.~T.~Grisaru and H.~N.~Pendleton,
  Nucl.\ Phys.\  B {\bf 124}, 81 (1977).
}

\lref\ParkePN{
  S.~J.~Parke and T.~R.~Taylor,
  Phys.\ Lett.\  B {\bf 157}, 81 (1985)
  [Erratum-ibid.\  {\bf 174B}, 465 (1986)].
}

\lref\BernUG{
  Z.~Bern, L.~J.~Dixon, D.~C.~Dunbar, M.~Perelstein and J.~S.~Rozowsky,
  Nucl.\ Phys.\  B {\bf 530}, 401 (1998)
  [arXiv:hep-th/9802162].
}

\lref\AnBern{
  C.~Anastasiou, Z.~Bern, L.~J.~Dixon and D.~A.~Kosower,
  Phys.\ Rev.\ Lett.\  {\bf 91}, 251602 (2003)
  [arXiv:hep-th/0309040].
}
\lref\KotikovFB{
  A.~V.~Kotikov, L.~N.~Lipatov and V.~N.~Velizhanin,
  Phys.\ Lett.\  B {\bf 557}, 114 (2003)
  [arXiv:hep-ph/0301021].
}

\lref\KotikovER{
  A.~V.~Kotikov, L.~N.~Lipatov, A.~I.~Onishchenko and V.~N.~Velizhanin,
  Phys.\ Lett.\  B {\bf 595}, 521 (2004)
  [Erratum-ibid.\  B {\bf 632}, 754 (2006)]
  [arXiv:hep-th/0404092].
}

\lref\BennaND{
  M.~K.~Benna, S.~Benvenuti, I.~R.~Klebanov and A.~Scardicchio,
  Phys.\ Rev.\ Lett.\  {\bf 98}, 131603 (2007)
  [arXiv:hep-th/0611135].
}

\lref\AldayQF{
  L.~F.~Alday, G.~Arutyunov, M.~K.~Benna, B.~Eden and I.~R.~Klebanov,
  arXiv:hep-th/0702028.
}


\Title{\vbox{\baselineskip12pt \hbox{SPIN-07/16} \hbox{
ITP-UU-07/24} }} {\vbox{\centerline{ Gluon scattering amplitudes  }
\centerline{ at strong coupling } }}
\bigskip
\centerline{ Luis F. Alday$^{a,b}$ and Juan Maldacena$^b$}
\bigskip
\centerline{\it $^a$Institute for Theoretical Physics and Spinoza
Institute} \centerline{Utrecht University, 3508 TD Utrecht, The
Netherlands}

\centerline{ \it  $^b$School of Natural Sciences, Institute for
Advanced Study} \centerline{\it Princeton, NJ 08540, USA}

\vskip .3in \noindent
 We describe how to compute planar gluon scattering amplitudes at strong coupling
 in ${\cal N}=4$ super Yang Mills by using the gauge/string duality.
The computation boils down to finding a certain classical string configuration whose boundary
conditions are determined by the gluon momenta. The results are infrared divergent. We introduce
the gravity version of dimensional regularization to define finite quantities. The leading
and subleading IR divergencies
are  characterized by two functions of the coupling that  we compute at strong coupling.
We compute also the full finite form for the four point amplitude and we find agreement
with a recent ansatz by Bern, Dixon and Smirnov.


 \Date{ }

\newsec{Introduction}
\noindent

In this article we describe a method for computing gluon
scattering amplitudes at strong coupling in ${\cal N}=4$ super
Yang Mills. Of course, these  amplitudes are infrared divergent
and are not good observables. Nevertheless,  in practical
computations for collider physics it is often useful to compute
these amplitudes as an intermediate step towards computing actual
well defined observables. Proper observables are IR finite (IR
safe), see {\it e.g} \refs{\StermanWJ,\BrockSZ,\EllisQJ}.
 Furthermore, in QCD computations it has proven useful to know the
${\cal N}=4$ super Yang Mills result as a building block \BernKQ.
Thus, it
is interesting to understand the behavior of scattering
 amplitudes at strong coupling. We perform the strong coupling
computation by using the gauge theory/gravity duality that relates
${\cal N}=4$ super Yang Mills to string theory on $AdS_5 \times
S^5$ \adscft . We consider planar amplitudes. On the string theory
side, the leading order result at strong coupling is given by a
single classical string configuration associated to the scattering
process. A similar result was  found by Gross and Mende
\grossmende\ in their investigation of fixed angle, high energy
scattering of strings in flat space. As we explain below,
the string theory scattering in $AdS$ is happening
at fixed angles and large energy and it is thus determined by a
classical solution. The final form for the color ordered planar
scattering amplitude of $n$ gluons at strong coupling is of the
form \eqn\scattem{ {\cal A} \sim  e^{i
  S_{cl} } = e^{ -{ \sqrt{\lambda } \over2 \pi } ({\rm Area})_{cl} } }
   where $S_{cl}$
denotes the classical action of a classical solution of the string
worldsheet equations, which is proportional to
  the area of the string world-sheet.
The solution depends on the momenta, $k^\mu_i$, of the gluons. The
whole dependence of the coupling is in the overall factor. Of
course, we expect $1/\sqrt{\lambda}$ corrections which we do not
compute. The structure of the IR divergences is precisely as
expected in the field theory. This comparison enables us to compute
the strong coupling expression for the function ${\cal
G}_0(\lambda)$ \refs{\bernetal,\MagneaZB,\StermanQN,\CataniBH}
characterizing the subleading IR divergent terms. The IR
regularization is done via dimensional regularization, as in the
field theory. This is achieved by using the gauge theory/gravity
duality for Dp-branes for general $p$ and then  performing an
analytic continuation in $p = 3 - 2 \epsilon$.

One of the motivations of this work was the very interesting  conjecture by
Bern, Dixon and Smirnov \bernetal\ (see also \AnBern ) for the all order form of the
$n$ gluon  MHV
scattering amplitudes. We have computed explicitly the full strong
coupling answer for the four point amplitude and found precise
agreement with their conjecture.
 Their conjecture is that the four point amplitude has the form
\eqn\conjecfop{ {\cal A}_4 = {\cal A}_{4}^{\text{tree}}
\exp\left[({\rm IR~divergent}) + {f(\lambda) \over 8}  (\log (s/t))^2 +
({\rm constant}) \right] } where $s,~t$ are Mandelstam variables,
$f(\lambda)$ is directly related with the cusp anomalous dimension
and the IR divergent terms are well characterized Sudakov-like
factors \bernetal .

Scattering amplitudes via the $AdS/CFT$ duality have been study in
many articles. See for example
 \refs{\RhoJM,\JanikZK,\polchstrass,\BrodskyPX,\AndreevVU,\BrowerEA} and references therein.

This paper is organized as follows. In section 2 we motivate and
describe
 the general prescription for
computing the leading order approximation for the amplitudes. In
section 3 we compute explicitly the classical string solution that
describes the four point amplitude, then we proceed with a discussion
of the structure of infrared divergencies and perform the comparison
with the results in \bernetal. Some conclusions and open problems
are sketched in section 4. Many additional remarks and technical
details are included in several appendices.

\newsec{Gluon scattering amplitudes}

Since scattering amplitudes of colored objects are not well
defined in the conformal theory it is necessary to introduce an
infrared regulator. The answer we obtain will depend on the
regularization scheme. Once we compute a well defined (IR
safe)  physical observable the IR regulator will drop out. An
example of a well defined physical observable is the amplitude for
a  process involving narrow jets going in some specific angular
directions. The answer will depend on the precise definition of
the jet observable, but not on the IR regulator. Further
discussion of these issues can be found in \refs{\StermanWJ,\BrockSZ,\EllisQJ}.
 A popular regularization scheme is dimensional regularization and we will
 use it in the next section. For the time being, it is convenient
to use a different IR regulator. In terms of the gravity dual,
this IR regulator is a D-brane that extends along the worldvolume
directions but is localized in the radial direction. In other
words,
 we start with the $AdS_5$ metric \eqn\adsfime{ ds^2 =R^2 \left[ {
dx_{3+1}^2 + dz^2 \over z^2 } \right] } and we place a D-brane at a
large value $z_{IR}$. In terms of the field theory,   such D-branes
arise, for example, if we go to the Coulomb branch of the theory.
 The asymptotic states are open strings that end on the
D-brane\foot{The color structure of the amplitude is not very
apparent with this regularization. We can get a glimpse of it by
introducing $N$ branes in the IR (instead of just one). In other
words, we go to a generic point in the Coulomb branch.  Then the
external gluons are strings stretching between these branes. The
color ordered amplitude would arise when we consider a configuration
where one open string ends on the brane where the next open string
starts. Namely, the $i$th open string goes between branes $i$ and
$i+1$.}. We then scatter these open strings. We are interested in
keeping the momentum fixed as we take away the IR cutoff.
 This means that the proper momentum of the strings living at $z_{IR}$
is very large, $ k z_{IR} \gg 1 $, where $k$ is the momentum in
field theory units ($k$ is conjugate to translations in $x$).
Thus, we are studying the scattering of open strings at fixed
angles and very high momentum. Such amplitudes were studied in
flat space by Gross and Mende,  \grossmende . The important
feature noted in \grossmende , is that amplitudes at high momentum
transfer are dominated by a saddle point of the classical
action\foot{ Here we concentrate on genus zero amplitude,
\grossmende\ computed also higher genus amplitudes using the same
idea.}.
 Thus, in order to compute the amplitude we simply have to compute a solution of
the classical action.  In our case we need to consider a
classical
 string in $AdS$.

\ifig\disktw{  Order of the particles in the  diagram and definition of $s$ and
$t$ for the four point amplitude.} {\epsfxsize1in\epsfbox{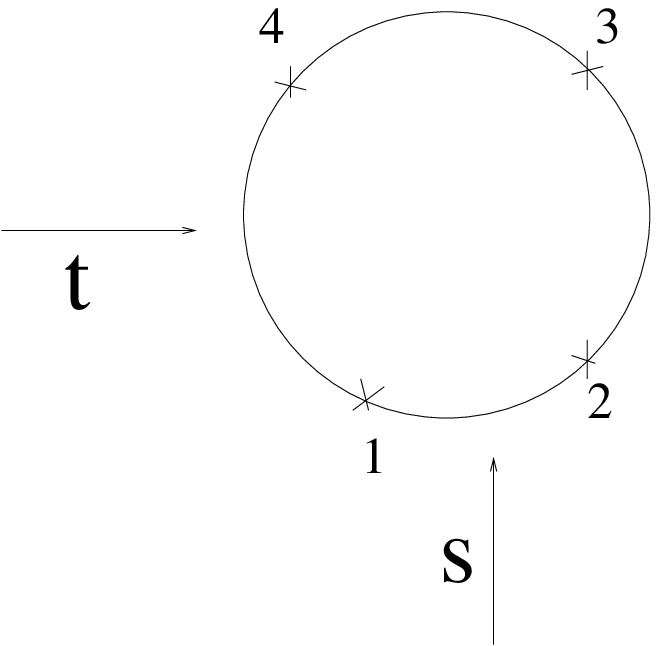}}

Let us first recall the flat space case \grossmende . We consider
scattering of open strings on a $D$ brane that is transverse to
the direction $z$, sitting at $z=z_{IR}$. We want to consider a
worldsheet with the topology of disk with vertex operator
insertions on its boundary. See figure \disktw .
 Near each vertex operator insertion
the solution should behave as \foot{We set $\alpha'=1$.}
\eqn\boundcon{ x^\mu \sim i k^\mu \log |w|^2 } where $k^\mu$ is
the momentum of the open string. In addition we require $z=z_{IR}$
on the boundaries of the worldsheet.   The solution is
\eqn\solutf{ x^\mu = i \sum_i k_i^\mu \log |w-w_i|^2 ~,~~~~~~~~~~z
= z_{IR} } Actually, we still need   to determine $w_i$. The
actual solution that minimizes the action has fixed values of
$w_i$, up to conformal transformations on the worldsheet. For
example, in the case of the four point function we can set three
of the $w_i$ to $w_1=0, ~w_3=1,~w_4=\infty$ and then
 $w_2$ is determined by inserting the above solution into the action and minimizing
with respect to $w_2$ which gives \eqn\valztwo{ w_2 = { s \over s
+ t }~ ; ~~~~~~~~~s = -(k_1+k_2)^2 ~,~~~~~~~~~t = -(k_1 + k_4)^2 }
The value of the action is then \eqn\classcat{ S_{flat}(s,t) = s
\log (-s) + t \log(- t) - (s+t) \log(- s-t) } and the leading
approximation to the scattering amplitude is then ${\cal A} =
e^{-S}$. Notice that the worldsheet has Euclidean signature in the
regime that $s,t < 0$ (spacelike $s$ and $t$ channel momentum
transfer, and timelike $u$ channel momentum transfer). In this
regime the solution \solutf\ is such that the coordinates along
the brane, $x^\mu$,  are purely imaginary.
 This amplitude is exponentially suppressed and
it represents the very small probability process where the two
incoming string states tunnel to the two outgoing ones. Notice
that we have an open string amplitude and the order of the
boundary vertex operators is important. Note that this leading
exponential behavior \classcat\ is independent of the particular string
states we are scattering, as long as we keep them fixed when we
increase the momentum transfer.

Let us now consider the $AdS$ case. Now it will be more difficult
to find the classical solution. However, there is one important
aspect of the classical solution that we can understand in a
qualitative way. Namely, we expect that the solution will be such
that in the central region of the collision the value of $z$ will
be much smaller than $z_{IR}$ and that it will be roughly
proportional to the inverse of $\sqrt{-s}$ or $\sqrt{-t}$
 since they
set the off shell momentum transfer of the process. This
expectation is based on the idea that the $z$ direction should
correspond to the off shell energy scale of the process. In
\polchstrass\ a similar scattering process was considered and it
was also found that the leading contribution came from $1/z \sim
\sqrt{(-s)}$.~ \foot{
  The main difference with our configuration  is that in \polchstrass\ the
   asymptotic states were
closed strings in the bulk. Thus they could move in the $z$
direction more easily than the open strings we consider, which are
attached to a D-brane at $z_{IR}$. The open  strings can move in
$z$ by stretching away from the D-brane, which is a stringy
excitation.}

  In order to state most simply the boundary conditions for the worldsheet it
is convenient to describe the solution in terms of T-dual
coordinates $y^\mu$ defined in the following way. We start with a
metric that contains \eqn\tuda{ ds^2 = w^2(z) dx_\mu dx^\mu + \cdots
} where $w$ is the warp factor. We define T-dual variables $y^\mu$
by \eqn\tudualva{
 \partial_\alpha y^\mu = i w^2(z) \epsilon_{\alpha \beta } \partial_\beta x^\mu
} In the regime under  consideration  the T-dual coordinates are
real and the worldsheet is Euclidean.
 In addition, the boundary condition
for the original coordinates $x^\mu$, which is that they carry
momentum $k^\mu$, translates into the condition that $y^\mu$ has
``winding'' \eqn\wind{ \Delta y^\mu = 2 \pi  k^\mu } Note that we
are not taking  the coordinates to be  compact (specially time!).
One can view this as purely a mathematical operation that makes it
easier to find the classical solutions.\foot{ One could consider
situations with compact spatial coordinates. Then we would be  doing
an honest T-duality on the spatial coordinates.} The T-duality will
also produce a non-trivial dilaton field, but it will not affect our
classical solutions. The T-dual  metric  is again $AdS_5$ after
defining $r = {R^2\over z} $ ~\foot{Notice that we do not do a
T-duality in the $z$ direction.} \eqn\tdualme{
 d \tilde s^2 = R^2 \left[ { dy_\mu dy^\mu
  + dr^2 \over r^2 } \right]  ~,~~~~~~~~~~~r={R^2 \over z}
 }
 The full Green Schwartz string action was computed in these
 ``T-dual'' variables in \kallosh .
\ifig\sixlines{ The kinematic data, namely the sequence of momenta
$k_1^\mu , \cdots, k_n^\mu$, translates into a sequence of
lightlike segments joining points in the T-dual space parametrized
by $y^\mu$. These points are separated by $2\pi k^\mu_i $. This
curve lives at $r=0$ in the T-dual AdS space \tdualme .  We need
to find a minimal surface that ends on this curve. }
{\epsfxsize1.5in\epsfbox{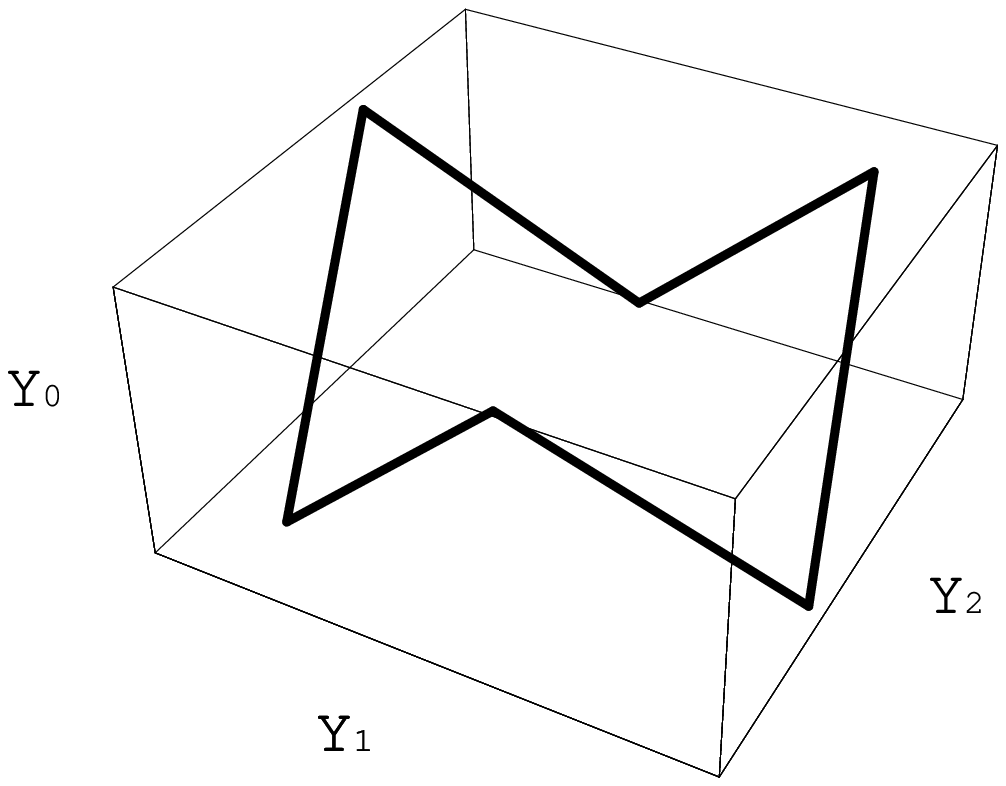}}

The solution we want is a surface which at  $r=R^2/z_{IR}$  ends
on a particular one dimensional line which is constructed as
follows. For each particle we have a lightlike segment joining two
points separated by \wind .
 We concatenate
these segments according to the  ordering of the open strings in
the disk diagram. This ordering is interpreted as the particular
color ordering of the amplitude. See \sixlines . The resulting
line consists of lightlike segments. The condition that the line
closes corresponds to the momentum conservation condition. As we
explained above the solution lives at values of $r > r_{IR}
=R^2/z_{IR}$. As we take the limit $z_{IR} \to \infty $ we find
that the boundary of the worldsheet moves to the boundary of the
T-dual metric \tdualme\ which is at $r=0$.
 From the point of view of the T-dual metric \tdualme\ the computation
 that we are doing is formally the
 same as the one we would do \wilsonloop\ if we
  were computing (in the classical string approximation)
 the expectation value of a Wilson loop
 given by a sequence of lightlike segments\foot{This prescription is vaguely
 reminiscent to the one used for non-commutative theories \grossetal .}.

In conclusion, the leading exponential behavior of the $n$-point
scattering amplitude is given by the area, $A$,  of the minimal
 surface that ends on a sequence of lightlike segments on the boundary
of \tdualme \eqn\amplitres{ {\cal A} \sim e^{ - {  R^2 \over 2 \pi
} A } = e^{ - { \sqrt{\lambda} \over 2 \pi } A } } The area
$A=A(k^\mu_1, \cdots,k_n^\mu)$ contains the kinematic information
about the momenta.
 The amplitude is color ordered in
a way that is reflected by the particular order in which the
lightlike segments are arranged along the boundary. There is no
information about the particular polarizations or states of the
gluons, which  contribute to
 prefactors in \amplitres , and
  are of subleading order in $1/\sqrt{\lambda}$.
  From the worldsheet point of view, we will need to supply this
information once we quantize around the classical solution. Such
terms are beyond the scope of this paper. Note that the $\lambda$
dependence is contained purely in the factor multiplying the
area\foot{ The area is computed in an $AdS$ space with radius one.
}.

The result \amplitres\ is still somewhat formal since the area is
infinite due to the infrared divergences that we mentioned above.
In order to find a finite answer we will need to regularize the
result. In the next section we will discuss the structure of the
infrared divergences and we will do an explicit computation for
the four point amplitude.

A reader familiar with AdS/CFT might be surprised by the fact that
in terms of the original coordinates we are setting boundary
conditions at $r=0$ or $z=\infty$ , rather than at $z=0$. This
apparent confusion goes away once we consider $AdS$ in global
coordinates, which is better in order to see whether we are at the
boundary or not. In those coordinates one can see that the surface
$z=\infty$   indeed intersects the boundary, as $x^\mu \to \infty$
and we will later check explicitly that the solution in terms of
the coordinates $(x^\mu,z)$ is intersecting the boundary of $AdS$.
This is discussed in more detail in appendix A.
 Thus, there is no contradiction with the general principle
stating that good observables are defined on the boundary of
$AdS$.

\newsec{Computation of the four point amplitude at strong coupling}

In this section we consider the planar four gluon amplitude. We
label the momenta as $k_1,~k_2,~k_3,~k_4$, where the subindex
indicates the color ordering, see \disktw . We consider the region
where particles 1 and 3 are incoming and particles 2 and 4
are outgoing. We label by $k$ the center of mass energy or
momentum of each of the incoming particles and we denote by
$\varphi$ the scattering angle in the center of mass frame.
We introduce the usual Mandelstam variables
\eqn\kininvar{\eqalign{
s & =- (k_1 + k_2)^2 = -2 k_1 .k_2 =- 4 k^2 \sin^2 {\varphi \over 2 }
\cr
t & = -(k_1 + k_4)^2 = -2 k_1.k_4= -4 k^2 \cos^2 {\varphi \over 2 }
\cr
u & = -(k_1 + k_3)^2 = -2 k_1.k_3 =  4 k^2 = -(s+t)
}}
We will focus on the region where $s,t < 0$  (they correspond to spacelike momentum transfer).
\ifig\fourlines{Sequence of lightlike segments which specifies the scattering
configuration. This figure lives at $r=0$ of the metric \tdualme .}
{\epsfxsize1.5in\epsfbox{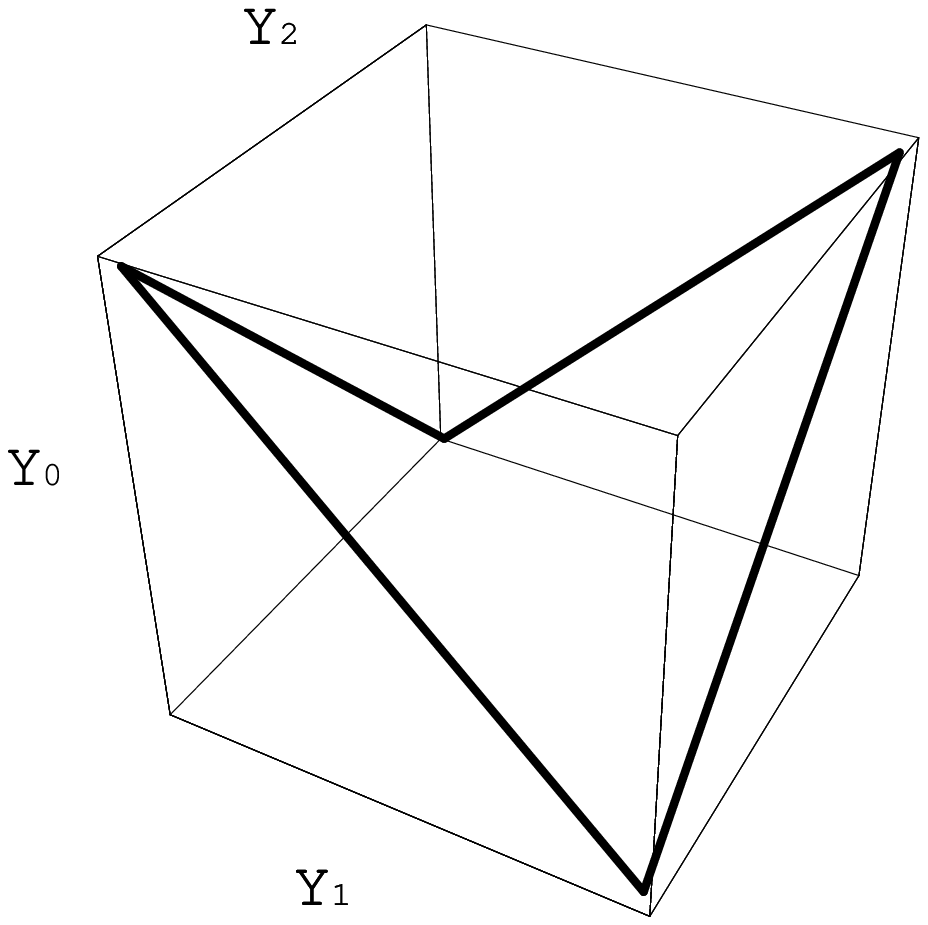}}
As we explained   above, we should find a classical string
solution specified by these momenta. It is simplest to think about the solution in
T-dual coordinates where the problem boils down to finding a minimal surface in
the T-dual
$AdS_5$ space \tdualme . This surface ends at $r=0$
 on a closed sequence of  lightlike segments whose sides are
specified by the lightlike momentum vectors $(2\pi)k^\mu_i$, see \fourlines .

\subsec{The lightlike cusp }

We start by considering  the solution near the cusp where two of the
lightlike lines meet. So we consider two semi infinite lightlike
lines meeting at a point. This case was considered in \kruczenski\
and it will prove useful for generating the solution we want.  It is
a surface that can be embedded in an $AdS_3$ subspace of $AdS_5$
\ifig\cusp{The lightlike cusp.  The thin lines shows the light
cone.} {\epsfxsize1.5in\epsfbox{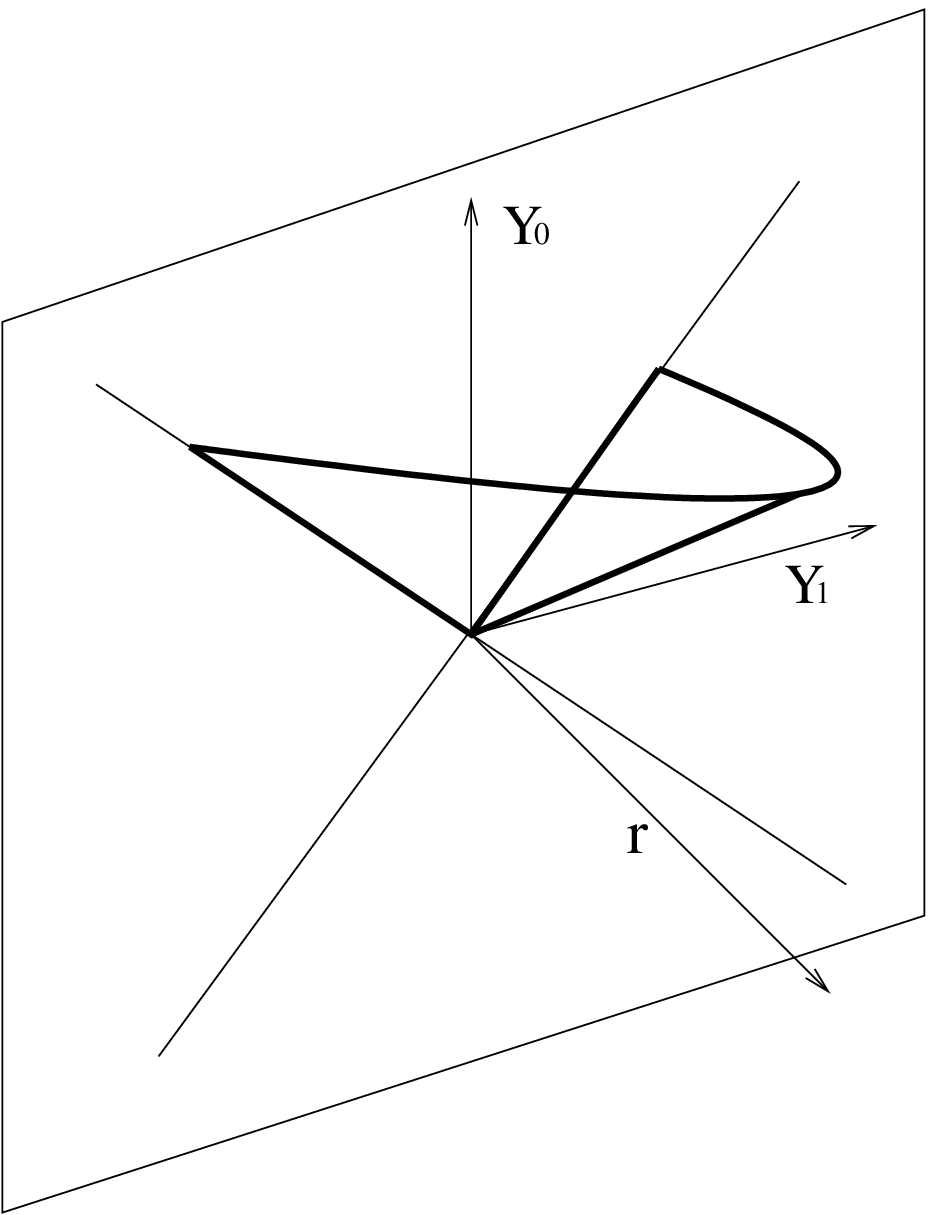}} \eqn\newmed{ ds^2 = { -
dy_0^2 + dy_1^2 + d r^2 \over r^{2} } } We are interested in
computing the surface ending on a light-like Wilson loop which is
along $y^1 = \pm y^0 $, $y^0 \geq 0$,\foot{One can also consider
Wilson loops along $y^0=\pm y^1$, $y^1>0$, the basic difference with
the ones considered here is that their world-sheet is Lorentzian and
$z$ is imaginary.} see \cusp . The problem has a boost and scaling
symmetry that becomes explicit if we choose the following
parametrization \eqn\cuspparam{y_0=e^\tau \cosh \sigma,~~~y_1=e^\tau
\sinh \sigma,~~~r=e^\tau w} Boosts and scaling transformations are
simply shifts of $\sigma$ and $\tau$. Then the Nambu-Goto action
 becomes \eqn\ngcusp{S ={R^2 \over 2 \pi } \left(\int d\sigma
\right)\int d \tau \frac{\sqrt{1- (w(\tau) +
w'(\tau))^2}}{w(\tau)^2}} One can explicitly check that
$w(\tau)=\sqrt{2} $ solves the equations of motion, hence the
surface is given by
\eqn\cuspsurface{r=\sqrt{2}\sqrt{y_0^2-y_1^2}=\sqrt{2} \sqrt{y^+
y^-} ~,~~~~~~~~~~~y^\pm = y^0 \pm y^1 } When we insert the solution
in the action \ngcusp\ we find that the lagrangian is purely
imaginary, this means that the amplitude ${\cal A} \sim e^{i S} $
will have an exponential suppression factor \eqn\suppref{ iS = - S_E
= - { R^2 \over 4 \pi } \int d\sigma d\tau } where $S$ is
the action for a spacelike surface embedded in the Lorentzian target
space that we are considering. This integral is infinite. We will
later discuss its regularization.

It is instructive to study this solution in terms of embedding
coordinates. These are coordinates where we view $AdS_5$ as the following
surface embedded  in $R^{2,4}$
\eqn\adsfc{
 - Y_{-1}^2  - Y_0^2 + Y_1^2 + Y_2^2 + Y_3^2 +  Y_4^2= -1
 }
The relation between these and the Poincare coordinates in \tdualme\ is
\eqn\intermofpu{\eqalign{ Y^\mu & = { y^\mu \over r} ~,~~\mu =0,\cdots ,3 ~ \cr
Y_{-1} + Y_4 & = { 1 \over r } ~,~~~~~~~~Y_{-1} - Y_4 = { r^2 + y_\mu y^\mu \over r }
}}
We can now write the surface corresponding to the cusp in terms of the equations
 \kruczenski\
\eqn\cuspemb{Y_0^2-Y_{-1}^2=Y_{1}^2-Y_4^2,~~~~~~Y_2=Y_3=0 }

\subsec{The four lighlike segments solution}

We now consider a Wilson loop containing four
light-like edges, which contains four cusps like  the one considered
above.
The configuration of lightlike
lines is shown in \fourlines .

In order to write the Nambu-Goto action it is convenient to
consider Poincare coordinates $(r,y_0,y_1,y_2)$, setting $y_3=0$,
 and parametrize the
surface by its projection to the $(y_1,y_2)$ plane. We consider
first the case with $s=t$ where the projection of the Wilson lines
in \fourlines\ is a square. The Nambu-Goto action is then the action
for two fields $y_0$ and $r$ living on a square parametrized by
$y_1$ and $y_2$.  The action reads \eqn\ngedge{ S = {R^2 \over2 \pi
} \int dy_1 dy_2 { \sqrt{ 1 + ( \partial_i r)^2 - (
\partial_i y_0)^2 - (
\partial_1 r \partial_2y_0 - \partial_2 r \partial_1 y_0 )^2 } \over
r^{ 2 } } }
By scale invariance, we
can change the size of the square.
We choose the edges of the square to be at $y_1,y_2=\pm 1$. The
boundary conditions are then given by
\eqn\squarebc{ r(\pm 1,y_2)=r(y_1,\pm 1)=0,~~~~y_0(\pm 1,y_2)=\pm
y_2,~~~y_0(y_1,\pm 1)=\pm y_1}
From the solution for the single cusp, we can obtain, after boost transformations,
the form of the solution in the vicinity of any of the cusps.
 The following expression for the fields can be easily seen to have
 the right behavior close to the cusps
\eqn\squaresol{y_0(y_1,y_2)=y_1 y_2,~~~~~r(y_1,y_2)=\sqrt{(1-y_1^2)(1-y_2^2)}
}
Remarkably it turns out to be a solution of the equations of motion.
When expressed in terms of embedding coordinates, the surface
is given by the equations
\eqn\equatsub{
Y_3=0 ~,~~~~~~Y_4=0 ~,~~~~~~Y_0 Y_{-1} = Y_1 Y_2
}
In fact this solution is related by the $AdS$ isometries ($SO(2,4)$ transformations)
 to
 the cusp solution \cuspemb \foot{In order to go from \cuspemb\ to
\equatsub\ we take $Y_2 \to Y_4$, $Y_0 \to { 1 \over \sqrt{2}}(Y_0 + Y_{-1} ) $,
$Y_{-1} \to { 1 \over \sqrt{2}}(Y_0 - Y_{-1} )$, $Y_1 \to  { 1 \over \sqrt{2}}(Y_1 + Y_{2} )$
and $Y_4 \to  { 1 \over \sqrt{2}}(Y_1 - Y_{2} )$}. The reader might be puzzled by
the following. We seem to have mapped a solution with two lightlike lines on the boundary
to one with four lightlike lines. The $AdS$ isometries are conformal transformations
on the boundary and  conformal transformations preserve
angles and cannot produce cusps where there were none.
The solution to this apparent puzzle is
that the cusp solution \cuspemb\ really has four cusps once it
is embedded in global coordinates
\kruczenski . We miss some of the cusps when we use Poincare coordinates
because those cusps  are on the
 boundary of the Minkowski space
parametrized by $y^\mu$. In fact this is a simpler alternative way
to derive the solution. Namely we start with the cusp solution
\cuspemb , notice that it really has four cusps and then map it
through conformal transformations to the solution we really want
\squaresol -\equatsub.

After we understood this point it becomes a simple exercise to compute the solution for
general $s$ and $t$. We simply need to apply an $SO(2,4)$
 transformation to the solution
we already have. Starting from \equatsub\ we perform a boost in the 04 plane and obtain
\eqn\afterboo{
Y_4 - v Y_0 =0 ~,~~~~~~~~~~~Y_{-1} \gamma (Y_0 - v Y_4)=
 \gamma^{-1} Y_0  Y_{-1} = Y_1Y_2  ~,~~~~~~~~~Y_3=0
}
where $\gamma^{-1}=\sqrt{1-v^2}$.

Let us now write the solutions in terms of worldsheet coordinates in conformal gauge.
Let us first go back to the solution for the case with $s=t$,  \squaresol ,
  and compute the induced metric on the worldsheet.
We find
\eqn\inducem{\eqalign{
ds^2 = &
 { dy_1^2 \over (1-y_1^2)^2 } + { dy_2^2 \over (1-y_2^2 )^2 } = du_1^2 + du_2^2
~,~~~~~~~~~{\rm where}~~~~~y_i = \tanh u_i
 }}
Notice that the metric on the worldsheet is Euclidean. More precisely, we
have a spacelike surface embedded in a Lorentzian target space.
Written in terms of $u_i$ coordinates  the solution \squaresol\ becomes
\eqn\soluvco{\eqalign{
y_1 &= \tanh u_1 ~,~~~~~~y_2 = \tanh u_2 ~,~~~~~~~r = { 1 \over \cosh u_1 \cosh u_2
 } ~,~~~~~~y_0 = \tanh u_1 \tanh u_2
\cr
Y_0 & = \sinh u_1 \sinh u_2 ~,~~~~~~Y_1 = \sinh u_1 \cosh u_2 ~,~~~~~~~~
Y_2 = \cosh u_1 \sinh u_2 ~,~~~~~~
\cr
Y_{-1} & =  \cosh u_1 \cosh u_2 ~,~~~~~~~Y_4=Y_3=0
} }
This is now a solution of the equations in conformal gauge, whose action reads
\eqn\confact{iS=-{R^2 \over 2 \pi } \int {\cal L} =-{R^2 \over 2 \pi }
\int du_1 du_2  { 1 \over 2 }
 { \left( \partial r\partial r +  \partial y_\mu \partial y^\mu
 \right)\over r^2 }}
 Note that the metric \inducem\ is Euclidean, this is responsible for the extra $i$ in
 this formula.
The lagrangian density evaluated on the solution is simply ${\cal L}=1$.
 Performing the boost \afterboo\ and a simple rescaling
 we now find the solution
for $s \not = t $ \eqn\soluvrom{\eqalign{r={a \over \cosh u_1 \cosh
u_2+b \sinh u_1 \sinh u_2},~~~~ y_0= {a \sqrt{1+b^2} \sinh u_1 \sinh
u_2 \over \cosh u_1 \cosh u_2+b \sinh u_1 \sinh u_2} \cr y_1={a
\sinh u_1 \cosh u_2 \over \cosh u_1 \cosh u_2+b \sinh u_1 \sinh
u_2},~~~~ y_2={a \cosh u_1 \sinh u_2 \over \cosh u_1 \cosh u_2+b
\sinh u_1 \sinh u_2} }} where $b = v \gamma$ and we consider $b<1$.
 The parameter $a$ sets the overall scale of the momentum.
  The solution approaches the boundary of
$AdS_5$ where $u_1$ or $u_2$ go to plus or minus infinity. These
four possibilities correspond to the four lightlike lines on the
boundary. For example, if we take $u_1 \to + \infty$ we find that
$r=0$ and \eqn\findre{
 y_1 = { a \over 1 + b \tanh u_2} ~,~~~~~~y_2 = {a \tanh u_2 \over 1 + b \tanh u_2 }, ~~~~
y_0 = a \sqrt{1 + b^2 } { \tanh u_2 \over 1 + b \tanh u_2 } } We see
that $y_1 + b y_2 =a$ and that we have a lightlike line going
between two points whose projections on the $y_1,y_2$ plane are
located at \eqn\twoendp{ {\rm A:} ~~~~y_1 = y_2 = { a \over 1 + b} ~,~~~~~~{\rm
and} ~~~~~{\rm B: } ~~~~y_1 = - y_2 = { a \over 1 - b } }
 which are reached at $u_2 \to \pm \infty$.
 \ifig\rombo{Projection of the light like lines on the $y_1,y_2$ plane for (a) $s=t$
 and (b) $s\not = t$.
 The line also moves in the time direction with a slope such that we get a lightlike
 line. Points
 on opposite vertices sit at equal times. Time goes up and down as we move from segment to
 segment. } {\epsfxsize2.5in\epsfbox{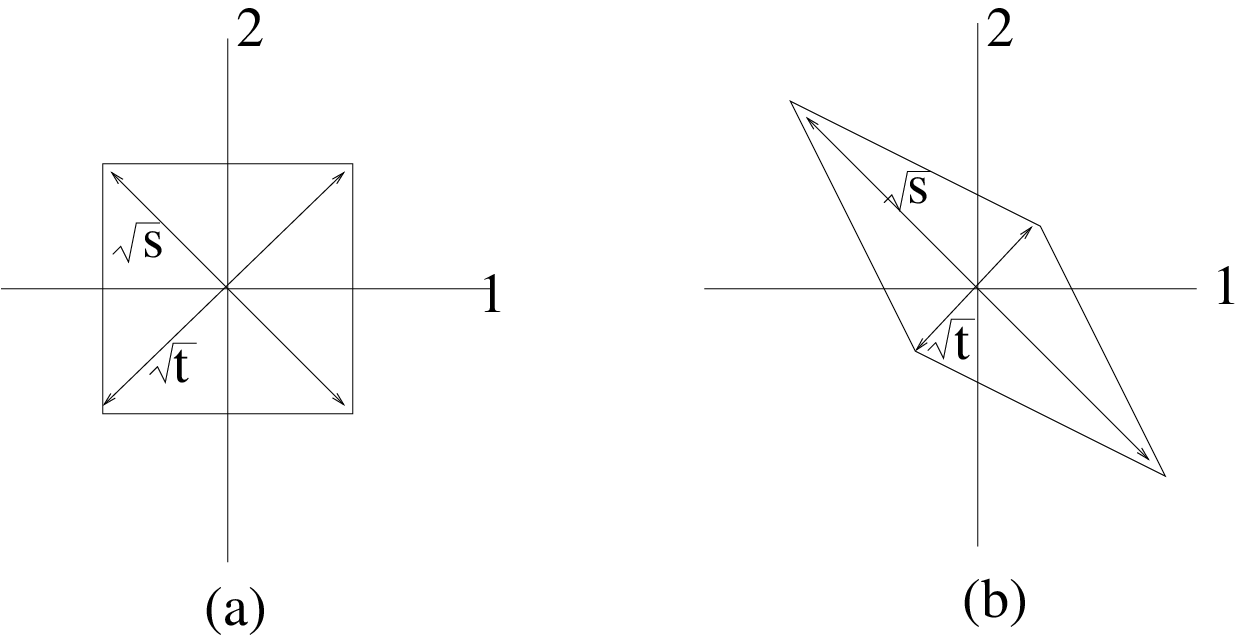}}
 By considering other
limits we get the other segments. The values of $s$ and $t$ are
given by the square of the distance between the vertex of two
non-adjacent cusps, see \rombo . In terms of the parameters $a$ and $b$ they are
given by
 \eqn\sandt{-s(2 \pi)^2 = {8 a^2 \over (1-b)^2},~~~~~-t (2 \pi)^2 ={8 a^2 \over (1+b)^2},
 ~~~~~{ s \over t } = { (1+b)^2 \over
(1-b)^2 } }
where the factors of $2\pi$ comes from \wind .

The solution and the value of the action are symmetric under $s
\leftrightarrow t$, which is a symmetry of the full problem.

It is also instructive to write the solution in terms of the
original $AdS$ coordinates \adsfime . We obtain
\eqn\origsol{\eqalign{ x_1 = & {i R^2 \over  a} \left( {u_2 \over
2}+{1 \over 4}\sinh 2u_2+b(-{u_1 \over 2}+{1 \over 4}\sinh 2
u_1)\right),~ \cr x_2 =  &{ i R^2 \over   a} \left( -{u_1 \over
2}-{1 \over 4}\sinh 2u_1+b({u_2 \over 2}-{1 \over 4}\sinh 2
u_2)\right)\cr x_0 =& {i R^2 \over 2 a }\sqrt{1+b^2}\left(\cosh^2
u_2-\cosh^2 u_1 \right) ~,~~~~~~ \cr z =&{  R^2 \over  a }
\left(\cosh u_1 \cosh u_2+b \sinh u_1 \sinh u_2 \right),}} We see
that for large $u_i$ the solution is such that it carries   momentum
in the spatial directions.   The solution lives in complexified AdS
space since it represents a sort of tunnelling solution. For $u_i
\to \pm \infty$
 the solution goes to the region where $z$ is large, which naively seems to corresponds to the
 IR of the field theory. On the other hand, we are also finding that $x^\mu$ are going to
 infinity at the same time. We can go to global coordinates and find that the solution
 \origsol\ indeed approaches the boundary of AdS. It touches the boundary on the surface
 with $1/z=0$ which corresponds to the region of the boundary that is the boundary of the
 Penrose diagram of the Minkowski slices parametrized by $x^\mu$.
 See appendix A.
In the central scattering region, $u_i \sim 0$, we have that $z
\sim R^2/a \sim R^2( { 1 \over \sqrt{-s} } + { 1 \over \sqrt{ - t
} } ) $ which is in agreement with the discussion in section two,
where we said that the scale of momentum transfer should set the
minimum value of $z$ that the solution explores \foot{The above
behavior can be qualitatively understood as the result of two
competing factors. On one side, high proper momentum transfers
suppress the scattering and this suppression can be lowered by
going to regions of smaller $z$. On the other hand we have to pay
a price for stretching the string away from $z=z_{IR}$ in the $z$
direction.  }.

We are now almost ready to evaluate the action. However, we should
note that there is a small subtlety. The action in terms of the
original coordinates and the action in terms of the T-dual
coordinates differ by a total derivative which will contribute with
a boundary term. The correct action to evaluate turns out to be the
area of the surface in the T-dual coordinates. When we consider the
problem in the original coordinates we should remember that we are
putting boundary conditions that fix the momentum at each boundary.
In order to have a proper variational principle we should add a
boundary term. This boundary term is precisely the one that turns
the original action into the T-dual action.

\subsec{Dimensional regularization in the gravity dual}

As we mentioned above the action of the solutions we have
discussed is infinite. A popular regularization method for ${\cal
N}=4$ super Yang Mills is the so called dimensional reduction
scheme \Siegel . In this scheme one goes to a general dimension $D
=4 - 2 \epsilon$ but one continues to use a theory with 16
supercharges. In other words, one considers the dimensional
reduction of ten dimensional super Yang Mills to $4 - 2 \epsilon$
dimensions. For integer dimensions these are precisely the low
energy theories living on Dp branes, where $p=D-1$. The gravity
dual of these theories involves the string frame metric \ItzhakiDD
\eqn\gravdual{\eqalign{ ds^2 = &  f^{-1/2} dx_{D}^2 + f^{1/2}
[dr^2  + r^2 d\Omega^2_{9-D}]~,~~~~~~~~~~~~~~D = 4 - 2 \epsilon
\cr f = & { c_D \lambda_D \over r^{ 8-D} } ~,~~~~~~~~~~~~~~~c_{D}
= 2^{4 \epsilon } \pi^{ 3\epsilon} \Gamma( 2 + \epsilon ) }} where
$\lambda_D = g^2_{D} N$ and $g_D^2$ is the coupling\foot{ This is
defined as the coefficient of the $D$ dimensional action in the
usual way $S = { 1 \over 4 g^2 } Tr[ F^2] $ and $Tr[T^aT^b] = { 1
\over 2 } \delta_{ab} $ where $T^a$ are the generators of SU(N).}.
We parametrize the coupling in $D$ dimensions in terms of the IR
cutoff scale $\mu$ as it was done in the field theory analysis of
\bernetal \eqn\coupldef{ \lambda_D = { \lambda \mu^{ 2 \epsilon}
\over (4 \pi e^{-\gamma} )^{\epsilon} } ~,~~~~~~~~ \gamma =
-\Gamma'(1) } where $\lambda = \lambda_4$ is the dimensionless
four dimensional coupling which is kept fixed as we vary
$\epsilon$.

From now on, we will drop the sphere part of the metric since it
does not play an important role. We now compute the metric for the
T-dual variables  defined through \tudualva\ (with $w^2
=f^{-1/2}$). We get \eqn\tdualreg{ ds^2 = f^{1/2} ( dy^2_D + dr^2)
= \sqrt{ c_D \lambda_D} \left( { dy^2_D + dr^2 \over r^{ 2 +
\epsilon} } \right) } Notice that the IR region of the original
metric corresponds to the region where $r \sim 0$ in the T-dual
metric \tdualreg .

In the region very close to $r\sim 0$ we cannot
trust the gravity dual and we should use the weakly coupled field
theory in the IR. In this field theory picture we can define the
asymptotic gluon states. We will see that we will only need the
geometry in the region where we can trust it if we are interested
in the strong coupling behavior.

\subsec{Evaluation of the action in dimensional regularization}

We regularize the theory by considering $Dp$-branes,
with $p=3-2 \epsilon$. We are then lead to the following action

\eqn\ngepsilon{ S  = { \sqrt{\lambda_D c_D }
\over 2 \pi }  \int { {\cal L}_{\epsilon=0} \over r^{\epsilon} }
}
where ${\cal L}_{\epsilon=0}$ is the lagrangian density for $AdS_5$, as in
\ngedge\ or \confact .

In order to understand how dimensional regularization works, let us perform the
computation for the lightlike cusp in general dimensions.
We still have the boost symmetry and we
 can make an ansatz similar to the one in \cuspparam\ but now the Lagrangian
depends explicitly on $\tau $
\eqn\newlagr{
S = { \sqrt{c_D \lambda_D} \over 2 \pi }  \int d\sigma
 \int d\tau e^{ - \epsilon \tau} { \sqrt{ 1 - ( w + w')^2 } \over w^{ 2 + \epsilon} }
}
It turns out that a constant $w$ is still a solution, but now the constant is
\eqn\cuspgen{
w= \sqrt{2}\sqrt{1 + {\epsilon \over 2} } ~~~\to ~~~~ r = \sqrt{2} \sqrt{ 1 + \epsilon/2}
\sqrt{ y^+ y^-}
} Inserting this solution into the action
and writing the integral as an integral over $y^\pm$ we get
\eqn\integirreg{
-i S =  A_\epsilon \int { dy^+ dy^- \over (2 y^+ y^-)^{1 + \epsilon/2} } =
{ 4 \over \epsilon^2 } { A_\epsilon \over (2 y_c^+ y_c^-)^{\epsilon/2} } ~,~~~~~~~
A_{\epsilon } = { \sqrt{ c_D \lambda_D} \sqrt{ 1 + \epsilon}
\over 8 \pi (1 + \epsilon/2)^{ 1 + \epsilon/2 } }
}
where $y_c^\pm$ is a cutoff for large $y_c^\pm$ \foot{In the computation of the actual amplitudes
this will be the momenta of the particles.}. We will later see that this double
pole is in agreement with field theory expectations.
One might be worried that the metric \tdualreg\ is getting highly curved as $r\to 0$
(for $\epsilon <0$). Fortunately this is not a problem if we are only interested
in determining the strong coupling behavior of the amplitude. We see this as follows.
The integral is cutoff at the point where
\eqn\regioncut{
 r^\epsilon \sim 1
 }
 But the effective curvature, ${\cal R}$, at that point is of the form
 \eqn\effecu{
 {\cal R} \sim { 1 \over \sqrt{\lambda } } r^{\epsilon } \ll 1  ~,~~~~~~~{\rm if} ~~~~
  ~~~\lambda \gg 1
 }
  Thus, for large lambda we can perform dimensional regularization
 without worrying about the region with strong curvature. If we wanted to understand the result
 at all values of $\lambda$, then it would be important to understand the whole region.
 A simple way to put it, is to say that dimensional regularization and strong
 coupling commute.

Let us now turn to the problem of the four-point scattering
amplitude. When $\epsilon \not =0$ we have a different lagrangian
\ngepsilon\ and we would need to find the solutions for the new
lagrangian, as we did above for the cusp. Fortunately, there is a
simple trick that allows us to find the solution to the accuracy
that we need\foot{ In principle, one could compute  the scattering
amplitudes for
 gluons in the case of $D=5,6$ if the center of mass energy is such that we
 can trust the corresponding gravity solutions \ItzhakiDD . That would
require  solving the equations for the new
lagrangian \ngepsilon\ with $\epsilon =- { 1 \over 2} ,-1$. }.
 We first note that if we have a lagrangian which has
the expansion ${\cal L} = {\cal L}_0 + \epsilon {\cal L}_1 +
\epsilon^2 {\cal L}_2 $, then we can expand the solutions of the
equations of motion as $q = q_0 + \epsilon q_1 + \cdots $. We will
be interested in evaluating the final answer to zeroth order in
$\epsilon$. However, since we have IR divergencies we find that the
leading order solution $q_0$ will give a leading double pole in
$\epsilon$. Thus when we do the formal expansion we mentioned above,
we will want to evaluate the action to the formal order
$\epsilon^2$, which in reality will be order $\epsilon^0$. Note that
then we will not need to know the solution to second order in
$\epsilon$ since that will contribute a term of the form $ \int {
\partial {\cal L}_0 \over
\partial q }|_{q_0} q_2 $
which vanishes due to the fact that $q_0$ obeys the zeroth order
equations of motion. For a similar reason the first order solution
$q_1$ will only contribute to terms of formal order $\epsilon^2$.
Thus to real orders $\epsilon^{-2}$ and $\epsilon^{-1}$ we can
simply evaluate the zeroth order solution in the new, $\epsilon$
dependent lagrangian and we will get an accurate enough answer.
Since the first order solution can only contribute to formal order
$\epsilon^2$ we will need an IR divergence of order $1/\epsilon^2$
to give a finite answer. These divergences only arise in the
cusps. Thus, we only need the $q_1$ solution near the cusps.
However, for the cusp region we know the solution \cuspgen . When
we have the general zeroth order solution of the problem, we can
get a solution that is accurate enough at the cusps by writing
\eqn\finalexp{r_\epsilon \sim \sqrt{1+{\epsilon \over
2}}~r_{\epsilon=0} ~,~~~~~~~~~~~ y^\mu_\epsilon \sim
y_{\epsilon=0}^\mu } where the $\epsilon =0$ solutions are the
ones we discussed above. Inserting these expressions into the
action \ngepsilon\ gives us an accurate enough answer to extract
the finite pieces in the amplitude \foot{In order to compute the regularized area we use the Nambu-Goto action in 
the static gauge chosen above, since the  corrected solution near the cusps \finalexp is expressed most simply in 
this gauge. }. A more detailed
analysis of the finite $\epsilon$ equations in the various regions
(near the cusps, near the lines) shows that the above argument is
indeed correct.

By inserting these expression into the action we get \eqn\fullS{
-i S= B_\epsilon \int_{-\infty}^\infty du_1 d u_2 ( \cosh u_1
\cosh u_2 + b \sinh u_1 \sinh u_2)^{\epsilon }\left(1+\epsilon
I_1+\epsilon^2 (I_2-2I_2^2)+... \right)} where
\eqn\terms{\eqalign{ I_1=&{(b^2-1)(\cosh 2u_1+\cosh 2u_2)-2(1+b^2)
\over 8 (\cosh u_1 \cosh u_2 + b \sinh u_1 \sinh u_2)^2} \cr
I_2=&{1+b^2-(1+b^2)\cosh 2u_1 \cosh 2u_2-2b \sinh 2u_1 \sinh 2u_2
\over 16 (\cosh u_1 \cosh u_2 + b \sinh u_1 \sinh u_2)^2} \cr
B_\epsilon = & { \sqrt{ \lambda_D c_D} \over 2 \pi } { 1 \over
a^\epsilon } }} where we have expanded up to terms that give
finite order answers in the final result. The integrals can be
performed   as explained in appendix B. The final result is
\eqn\fina{i S=- B_\epsilon \left({\pi \Gamma[-{\epsilon \over
2}]^2 \over \Gamma[{1-\epsilon \over 2}]^2} ~_2F_1({1 \over
2},-{\epsilon \over 2},{1-\epsilon \over 2};b^2) +1/2 \right)}
%
%
It is then straightforward to expand in powers of $\epsilon$ up to
finite contributions. We need to recall the expressions for $s,t$ in
 \sandt\ and also the formulas for $c_D$   \gravdual\ and $\lambda_D$ \coupldef .
 Putting all this together we get the final answer
\eqn\finalans{{ \cal A} = e^{i S} = \exp \left[ iS_{div} +
{\sqrt{\lambda }  \over 8 \pi  } \left( \log{ s \over t} \right)^2
+ \tilde C \right] } \eqn\localdiv{\eqalign{ \tilde C = & {
\sqrt{\lambda } \over 4 \pi } ({\pi^2 \over 3}+2\log 2-(\log 2)^2
) \cr iS_{div} = & 2 iS_{div,s} + 2 iS_{div,t} }} where
$S_{div,s}$ and $S_{div,t}$ are the divergent pieces associated to
each cusp or pair of consecutive gluons. There are two pairs with
total squared momentum $t$ and two with $s$. We have \foot{We have
started with a cutoff scale $\mu$ as done in the field theory
analisys of \bernetal
 , however, one can verify that if we have started with $\sqrt{2}\mu$ instead,
 all the terms containing $\log 2$ in our final result would disappear. }
 \eqn\pizzadiv{ i S_{div,s } = - { 1 \over \epsilon^2 } { 1 \over 2
\pi } { \sqrt{\lambda \mu^{2 \epsilon } \over(- s)^\epsilon }} - { 1
\over \epsilon} \frac{1}{4\pi}( 1 - \log 2) { \sqrt{\lambda \mu^{2
\epsilon } \over (-s)^\epsilon }} }
and $S_{div,t}$ is given by a similar expression with $s \to t$. We now compare what
we obtained here with the field theory results and conjectures in \bernetal .

\subsec{IR divergences of amplitudes }

In this section we recall some general results on IR divergences of
scattering amplitudes \refs{\bernetal,\MagneaZB,\StermanQN}.
We will focus here on planar diagrams and color ordered amplitudes.
The first result is that the IR singularities of the amplitude can
be associated to consecutive   gluons in the color ordered
amplitude. This is fairly clear once we recognize that the IR
singularities come from low momentum gluons and we use that we
consider only planar diagrams. The leading divergence goes like
\eqn\leadiv{ {\cal A} = e^{ - { f(\lambda) \over 4 }   (\log \mu )^2
} } where $\mu$ is a mass scale that is acting as an IR cutoff. We
have one factor like \leadiv\ for each pair of consecutive gluons.
The reason that the divergencies exponentiate in this way is the
following. The divergencies come from the exchange of soft gluons
among two consecutive hard gluons.
 In the limit that the hard gluon momenta are infinite,
the configuration of hard gluons is invariant under two symmetries:
boosts and scale transformations. Let us denote by $\sigma$ the
boost parameter and by $\tau$ the parameter generating scale
transformations, $x^\mu \to e^\tau x^\mu$. Thus naively the
amplitude will be of the form \eqn\amplif{ {\cal A}_{div} = e^{-
h(\lambda) \Delta\sigma \Delta\tau } } where $\Delta \sigma$ and
$\Delta \tau$ is the range of scale and conformal transformations
where the approximation of exact boost and scaling symmetry are
valid. (Recall the classical string
 result \suppref ). This is conceptually
similar to the problem of computing the partition function of a
system that is invariant under time translations and spatial
translations. The answer will be $Z = e^{ - f L T}$ where $L$ and
$T$ are two IR cutoffs. In fact, this can be made more explicit by
choosing coordinates in $R^{1+3}$ that lead to a metric which is
Weyl equivalent to a metric where these symmetries are
explicit\foot{ One can write the metric of $R^{1+3}$ as $ds^2 = e^{2
\tau} ( - d\tau^2  + ds^2_{H_3} )$. Since the theory is conformal
invariant we can drop the conformal factor in the metric. We can
further choose coordinates in hyperbolic space so that we have now
the metric $ds^2 = - d\tau^2 + d \rho^2 + \cosh^2 \rho d\sigma^2 +
\sinh^2 \rho d \varphi^2 $. Boosts and scale transformations
coorrespond now to shifts of $\tau$ and $\sigma$. }.
 The function $f$ that appears in \leadiv\
 is proportional to the cusp
anomalous dimension for a Wilson loop in the fundamental
representation \foot{For two Wilson lines forming a spacelike cusp where the
two lines differ by a large boost parameter $\gamma$ the anomalous dimension is
$\langle W \rangle \sim  e^{ - {f \over 4 } \gamma \log(L_{IR}/L_{UV} ) }$, where $L$ are
UV and IR cutoffs. We refer the reader to \refs{\kruczenski,\CollinsBT,\KorchemskySI}
 and references therein for a discussion of all these ideas.}.
 Of course we also need another scale so that the log
in \leadiv\ makes sense. This is provided by $-s= (p_1 +p_2)^2 = 2
p_1 p_2$ where $p_1$ and $p_2$ are the momenta of the two lines.
 The function
$f$ in \leadiv\ also appears when one computes the dimension of
operators of high spin, $S$, of the schematic form $Tr[ \Phi
\partial^S \Phi] $, with $\Phi$ in the adjoint \KorchemskySI . These operators have
twist \eqn\twistop{ \Delta - S =   f(\lambda ) \log S } This
function was computed perturbatively up to four loops in
\refs{\BernEW,\CachazoAZ} , at strong coupling using strings in
$AdS$ in \refs{\gkp,\FrolovAV} and given exactly as a solution of
an integral equation in \bes~ using integrability (see also
\integrothers ). Of course, in addition to the leading divergence
in \leadiv\ we can have a subleading divergence involving a single
log. Thus, we can introduce a second function $g$ through
\eqn\leadisubl{ {\cal A}_{div,s} = \exp\left\{ - {f(\lambda) \over
16 } \left( \log { \mu^2 \over (-s)} \right)^2  - { g(\lambda)
\over 4} \left( \log {\mu^2 \over (-s)} \right) \right\} } Where
we have defined a new function $g$. The precise form of $g$ will
depend on the precise definition of the IR regulator since
shifting the log in the first term by a constant can affect the
form of $g$. \foot{What we called $g$ here is called ${\cal G}_0$
in \bernetal .} In other words, changing $\mu \to \mu  \kappa$ we
change \eqn\chagde{ g \to g +  f \log \kappa }

Let us now review how these divergences appear when we perform
dimensional regularization. The double logs in \leadiv\ will arise
if we have a double pole in $\epsilon$. Thus, in dimensional
regularization the divergences should organize as
\refs{\bernetal,\MagneaZB,\StermanQN} \eqn\dimresg{ {\cal A}_{div,s}
= \exp\left\{ - { 1\over 8 \epsilon^2 } f^{(-2)}\left( { \lambda_4
\mu^{2 \epsilon } \over s^\epsilon} \right) - { 1 \over 4  \epsilon
} g^{(-1)}\left({ \lambda_4  \mu^{2 \epsilon } \over s^\epsilon}
\right) \right\} } This formula, together with \coupldef ,  gives a
precise definition for $g$. We now see that expanding in $\epsilon$
we reproduce the double logs in \leadiv\ if \eqn\equforf{
 \left(\lambda { d \over d \lambda } \right)^2 f^{(-2)}(\lambda ) = f(\lambda) ~,~~~~~~~~~
\lambda { d \over d \lambda }   g^{(-1)}(\lambda ) = g(\lambda) } By
comparing the general expression for the IR divergence \dimresg\ to
our result \pizzadiv\ we can compute the functions $f$ and $g$ at
strong coupling. We find \eqn\valgres{f=4  { \sqrt{\lambda } \over 4
\pi }  ,~~~~~~~~g = { \sqrt{\lambda } \over 4 \pi } 2 (1 - \log 2 )
}
Notice that the square root of $\lambda$ dependence introduces a factor of $2$ when we integrate
as in \equforf .
Of course, the function $f$ has the same value that was computed in other ways in
\refs{\gkp,\kruczenski} this is implied by the general theory of IR divergences  we have just
reviewed.
In appendix C we discuss the extrapolation of the weak coupling results to strong coupling
and the comparison with our answer \valgres .

\subsec{Field theory results for the four point amplitude}

The four point scattering amplitude in maximally supersymmetric
theories in $D$ dimensions has the form \eqn\maxsup{ {\cal A} =
{\cal A}_{Tree} a(s,t) } This   is completely determined by the fact
that the theory has 16 supersymmetries.\foot{See for instance
\refs{\GrisaruVM,\GrisaruPX,\ParkePN} and appendix E of \BernUG .} A
simple way to understand it is the following. In a theory with 16
supersymmetries the generic massive representation has $2^8$
components. The gluons live in a representation which has $16=2^4$
components. This is possible because it is  a massless
representation and half of the supercharges act trivially. When we
think about a $2\to 2$ scattering process we can see that the state
formed by two gluons transforms as a massive representation and has
$2^8$ states, which is the same as the total number of polarization
states for two gluons. This means that there is a unique
intermediate state. This fixes the $S$ matrix uniquely up to a
common function.\foot{ A similar argument was used in \beisertsnlin\
for $S$ matrices on spin chains. } Of course this $S$ matrix
determined by the symmetries will be equal to the tree level
S-matrix which preserves all symmetries. This explains the form of
the four point scattering amplitude in \maxsup\ in all dimensions.
In four dimensions the function $a(s,t)$ has to obey further
constraints. If we demand scaling symmetry we would conclude that it
is a function of $a(s/t)$ and if we further demand that it is
invariant under special conformal symmetries we would conclude that
$a$ is a constant. However, due to the IR divergencies we have an
additional dependence on the cutoff, which enables the scattering
amplitude to be a non-constant function of $ s,t$. \foot{It seems
possible that the functional form of the IR regularized result
 is still determined by the action of the
special conformal generators, though we were not able to prove it. }

Bern, Dixon and Smirnov \bernetal\ made a  conjecture for the exact
form of the four loop amplitude of the form \eqn\bdsconj{ {\cal A} =
{ \cal A}_{tree} ({\cal A}_{div,s})^2 ({\cal
A}_{div,t})^2\exp\left\{ { f(\lambda ) \over 8 } [  ( \log{ s\over
t} )^2 + 4 \pi^2/3 ] + C(\lambda) \right\} } where $  C(\lambda) $
is only a  function of the coupling,
 $f$ is the same function appearing above \leadiv ,
 and ${\cal A}_{div,s}$ is given by \dimresg .
We see that the momentum dependent
 finite piece of our strong coupling expression has precisely the form
predicted by \bernetal , including all numerical factors after we
include the appropriate strong coupling form for the cusp anomalous
dimension \valgres . Unfortunately we cannot test \bernetal\ for the
 the constant
finite pieces, though we could  do it if we
computed the $n$ point amplitudes, which are also predicted in
\bernetal .
Of course, what we can definitely say is that
 \localdiv\  implies a strong coupling value
for a combination of the constants introduced in \bernetal .

\newsec{Conclusions}

In this paper we have given a prescription for computing
planar gluon scattering amplitudes at strong coupling using the gauge
theory/gravity duality. The computation involves finding a
classical string solution moving in (complexified)
 $AdS_5$ with a prescribed asymptotic
behavior determined by the gluon momenta. The computation can be
regularized using the gravity version of dimensional regularization.
Though our prescription would work for an arbitrary $n$ point
amplitude, we have only computed explicit answers for the four point
amplitude. We found results that agree in all detail with the
conjecture of Bern, Dixon and Smirnov \bernetal . The structure of
IR divergences is precisely as expected from general field theory
reasoning. The detailed comparison enables us to extract the strong
coupling form of the function that determines the subleading
divergent terms. The leading divergent terms are determined by the
cusp anomalous dimension which is already known at strong coupling
\gkp \kruczenski .

It is amusing that the computation is mathematically similar to the
computation we would do in order to compute the expectation values
of locally lightlike Wilson loops. This formal similarity, however,
might disappear at higher orders in $1/\sqrt{\lambda}$.

It seems possible to try to check the conjecture of \bernetal\ for
$n$ point functions for $n>4$. Since the $AdS_5$ sigma model is
integrable, it should be possible to find the appropriate classical
solutions and evaluate the action. Or even evaluate the action
without finding the full explicit form of the classical solutions!

It would also be interesting to learn how to do computations of scattering amplitudes
exactly as a function of the coupling by using integrability and spin chains. For this
we note that performing computations on $S^3 \times R$ would be another way to impose an IR
cutoff. In that situation it would seem that a single gluon would correspond to a configuration
containing spiky strings, similar in spirit to the ones in \spiky . It might be possible to
consider the scattering of spikes by relating them to some objects on the spin chain.

Hopefully, these amplitudes might be helpful for understanding aspects of QCD and the
transition between the perturbative and non-perturbative regimes.

{\bf Acknowledgments}

We would like to especially thank Z. Bern for discussions,
explanations and comments on the draft. We would also like to thank
L. Dixon, R. Roiban, M. Spradlin and G. Sterman for discussions.

This work   was  supported in part by U.S.~Department of Energy
grant \#DE-FG02-90ER40542. The work of L.F.A was supported by VENI
grant 680-47-113.

\newsec{Appendix A: Solution in the original global coordinates}

We start by considering the solution in terms of the original
Poincare coordinates (we set $R=1$ as it doesn't play any role in
the following discussion.)\eqn\origsolp{\eqalign{ x_0 = &{i \over 2
a}\sqrt{1+b^2}\left(\cosh^2 u_2-\cosh^2 u_1 \right) ~,~~~~~~z ={1
\over a} \left(\cosh u_1 \cosh u_2+b \sinh u_1 \sinh u_2 \right),
\cr x_1 = & {i \over a} \left( {u_2 \over 2}+{1 \over 4}\sinh
2u_2+b(-{u_1 \over 2}+{1 \over 4}\sinh 2 u_1)\right),~ \cr x_2 = &
{i \over a} \left( -{u_1 \over 2}-{1 \over 4}\sinh 2u_1+b({u_2 \over
2}-{1 \over 4}\sinh 2 u_2)\right) }} In order to study the
intersection of our surface with the boundary of $AdS$ we write the
solution in term of embedding coordinates \eqn\intermofpu{\eqalign{
X^\mu & = { x^\mu \over z} ~,~~\mu =0,\cdots ,3 ~ \cr X_{-1} + X_4 &
= { 1 \over z } ~,~~~~~~~~X_{-1} - X_4 = { z^2 + x_\mu x^\mu \over z
}
 }}
The boundary of $AdS$ in global coordinates is then parametrized by
the space of coordinates satisfying
\eqn\bound{ -X_{-1}^2-X_{0}^2+X_{1}^2+...+X_{4}^2=0}
quotiented by overall rescalings, but they cannot all be zero.
The geometry of the boundary is $R \times S^3$.

As discussed in the main text, we impose boundary conditions at
$z=\infty$. Notice that $z=\infty$ implies the equation $X_{-1} +
X_4 =0$ which gives a lightlike surface on the bulk that intersects
the boundary also on a lightlike surface. The region on the boundary
that is bounded by this lightlike surface is conformal to Minkowski
space. In fact, that region corresponds to the Minkowski space
parametrized by $x^\mu$ at $z =0$. However, the $z=\infty$ surface
also extends in the interior of $AdS_5$. In order to see whether the
region $u_i \to \pm \infty $ is on the boundary or not we need to
check the behavior of the $x^\mu$ coordinates. Without loss of
generality lets assume $u_1 \gg u_2\sim 1$. In this regime we obtain
the following expression for the embedding coordinates\foot{We thank
Shesansu Pal for pointing out an error in the original version of
these equations.} \eqn\embbound{\eqalign{X_0=&-{i \sqrt{1+b^2} \over
4}{ e^{u_1} \over \cosh u_2+ b \sinh u_2} ,~~~~~X_1={i b \over 4}{
e^{u_1}\over \cosh u_2 + b \sinh u_2 },~~~~~X_3=0
\cr~~~~~~~~~~~~~~~~~~~&X_2=- {i \over 4}{ e^{u_1} \over \cosh u_2+ b
\sinh u_2},~~~~~~
 X_{-1}=-X_{4}=\frac{u_1(b^2-1)e^{u_1}}{8 a(\cosh u_2+b \sinh u_2)}}}
These coordinates are complex but large. Thus we can say that they live in the
complexified boundary, which is defined by \bound\ but now the coordinate are complex and
we allow complex rescalings.

\newsec{Appendix B: A useful integral.}

In this appendix we show how to compute the following integral
\eqn\somef{ S=\int_{-\infty}^\infty du d v ( \cosh u \cosh v + b
\sinh u \sinh v)^{\epsilon} } This integral can be done by expanding
the integrand as a power series on $b$, then integrating term by
term and finally performing the sum \eqn\regdim{S=\sum_{l=0}^\infty
{\Gamma[\epsilon+1] \over \Gamma[\epsilon+1-l] l!}b^{2l} \left( \int
du (\cosh u)^{\epsilon} (\tanh u)^l\right)^2 } The double integral
becomes a sum of integrals on a single variable
\eqn\regidentity{\int_{-\infty}^\infty du (\cosh u)^{\epsilon}
(\tanh u)^l ={(1+(-1)^l)\Gamma[{1+l \over 2}]\Gamma[-{\epsilon \over
2}] \over 2 \Gamma[{1+l-\epsilon \over 2 }]} } This identity is
valid when $\epsilon<0$. Finally, performing the sum we obtain
\eqn\simpleregdim{S={\pi \Gamma[-{\epsilon \over 2}]^2 \over
\Gamma[{1-\epsilon \over 2}]^2} ~_2F_1({ 1\over 2},-{\epsilon \over
2},{1-\epsilon \over 2};b^2) } This expression is valid for all
values of $\epsilon<0$. However, we are interested in the behavior
for small $\epsilon$. We find that \eqn\finrestu{F \equiv ~_2F_1({1
\over 2},-{\epsilon \over 2},{1-\epsilon \over 2};b^2)=1+{1 \over
2}\log(1-b^2)\epsilon + {1 \over 2} \log(1-b) \log(1+b)\epsilon^2
 + {\cal O}(\epsilon^3)
}
We find it convenient to express this expansion as follows
\eqn\estra{\eqalign{F &= F_s + F_t - { \epsilon^2 \over 4} \left(
\log{ 1 + b \over 1 -b } \right)^2 \cr F_s &= {1 \over 2} + { \epsilon \over
2} \log(1+b) + { \epsilon^2 \over 4} (\log(1+b ))^2 = { 1 \over 2 }
( 1 + b)^\epsilon,~~~~F_t= { 1 \over 2 } ( 1 - b)^\epsilon}}
Where all the equalities are valid up to order $\epsilon^2$.

 \newsec{Appendix C:  Extrapolating the weak couping results for $g(\lambda)$ to
strong coupling.}

In this appendix we give a rough estimate for $g(\lambda)$ at strong
coupling by using its weak coupling expansion\foot{
This calculation was suggested to us by Z. Bern, L. Dixon and R. Roiban , who also
did it. }. We follow closely the
idea proposed in \KotikovFB~in order to obtain an approximate
expression for the cusp anomalous dimension valid at all values of
the coupling constant.

The weak coupling expansion for the cusp anomalous dimension is
\KotikovER \eqn\weakcusp{f(\lambda)=8 \left({\lambda \over 16 \pi^2}
\right)-16 \zeta_2 \left({\lambda \over 16 \pi^2} \right)^2+176
\zeta_4 \left({\lambda \over 16 \pi^2} \right)^3+...} This
perturbative information together with its strong coupling behavior
$f(\lambda) \sim \sqrt{\lambda}$ suggest a relation of the form
\eqn\rel{ \left({\lambda \over 16 \pi^2}
\right)^n=\sum_{r=n}^{2n}C_r (f(\lambda))^r} For a given value of
$n$, once we fix the coefficients $C_r$ by using the perturbative
data, \rel\ predicts a given strong coupling limit. \foot{Actually,
see discussion in \BernEW, if one introduces enough perturbative
information, namely $n \geq 4$, the value predicted for $f(\lambda)$
at strong coupling is in very good agreement with the known strong
coupling value. Furthermore, combining the known weak and strong
coupling information one can get an answer accurate to 1 $\%$ (when
compared against the answer from the BES equation) for the whole
range of the coupling constant.}

At weak coupling $g(\lambda)$ was computed up to three loops
\bernetal \eqn\gexp{g(\lambda)=-4 \zeta_3 \left({\lambda \over 16
\pi^2} \right)^2+8(4 \zeta_5+10/3 \zeta_2 \zeta_3) \left({\lambda
\over 16 \pi^2} \right)^3+... } To repeat the analysis discussed
above in order to extract the strong coupling behavior of
$g(\lambda)$ is subtle due to two features. First, $g(\lambda)$
starts at two loops, so the relation \rel\ should be modified
accordingly. Second, note that if our prediction is correct,
$g(\lambda)$ should change its sign in going from weak to strong
coupling. This does not occur for $f$.

A possibility is to define a new function $\tilde{g}(\lambda)$,
differing from $g(\lambda)$ by a constant times $f(\lambda)$. As
seen in the main text, this shift corresponds simply to a change in
the IR regulator.

\eqn\shiftedg{\tilde{g}(\lambda)=g(\lambda)+\xi f(\lambda)}

For positive $\xi$, $\tilde{g}(\lambda)$ will be positive both at
weak and strong coupling. It is then straightforward to repeat the
analysis of \KotikovFB\ for $n=3$ (since we have three coefficients
at our disposal). The value obtained at strong coupling should be
then compared with our prediction

\eqn\gtstrong{\tilde{g}(\lambda)=\left(2(1-\log(2))+4 \xi \right)
{\sqrt{\lambda} \over 4 \pi}+...}

The following figure shows a comparison between the value predicted
by our computation \gtstrong\ (solid blue line), and the value
obtained by a naive extrapolation as explained above (dashed red
line), for the range $0<\xi<2$

\ifig\disktw{Comparison between the result for $\tilde{g}$ obtained
from our computation (solid blue line) and a naive extrapolation
from weak coupling (dashed red line).}
{\epsfxsize2.5in\epsfbox{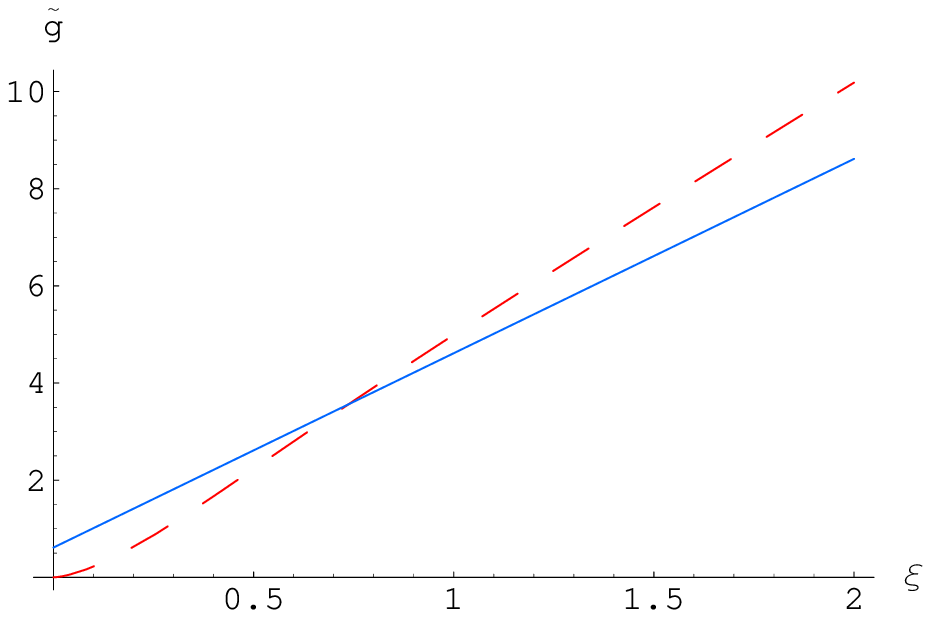}}

Note that for $\xi=0$ the assumption \rel\ is not valid. At the
particular value $\xi=\log(2)/2$ the strong coupling prediction is
particularly simple\foot{Note that if $\log(2)$ has
transcendentality 1, then such a shift doesn't break
transcendentality.} \eqn\gtstrong{\tilde{g}(\lambda)=2
{\sqrt{\lambda} \over 4 \pi}+...} whereas the value from the
extrapolation turns out to be \eqn\gtstrong{\tilde{g}(\lambda)
\approx 1.37 {\sqrt{\lambda} \over 4 \pi}+...} This differs by a $30
\%$ from the predicted number. All together, given the little that
is known for the function $g(\lambda)$, all that we can say is that
a naive extrapolation from weak to strong coupling seems to be
compatible with our results.

Of course, the above analysis is by no means rigorous. In order to
make a more precise comparison one would like to have the analog of
the BES equation from which $g(\lambda)$ could be computed at any
order, then one could proceed along the lines of
\refs{\BennaND,\AldayQF}.

\listrefs

\bye